\documentclass[10pt]{article}
\usepackage[dvips]{graphicx}

\usepackage{epsfig}
\usepackage{graphicx}
\usepackage{indentfirst}
\usepackage{amsmath}
\usepackage{amsfonts}
\usepackage{amssymb}

\begin{document}

\begin{center}
\begin{flushright}\begin{small}    UFES 2011
\end{small} \end{flushright} \vspace{1.5cm}
\huge{Static Anisotropic Solutions in $f(T)$ Theory} 
\end{center}

\begin{center}
{\small  M. Hamani Daouda $^{(a)}$}\footnote{E-mail address:
daoudah8@yahoo.fr}\ ,
{\small  Manuel E. Rodrigues $^{(a)}$}\footnote{E-mail
address: esialg@gmail.com}\ and
{\small    M. J. S. Houndjo $^{(b)}$}\footnote{E-mail address:
sthoundjo@yahoo.fr} \vskip 4mm

(a) \ Universidade Federal do Esp\'{\i}rito Santo \\
Centro de Ci\^{e}ncias
Exatas - Departamento de F\'{\i}sica\\
Av. Fernando Ferrari s/n - Campus de Goiabeiras\\ CEP29075-910 -
Vit\'{o}ria/ES, Brazil \\
(b) \ Instituto de F\'{i}sica, Universidade Federal da Bahia, 40210-340, Salvador, BA, Brazil\\
\vskip 2mm

\end{center}

\begin{abstract}
In a previously work, we undertook a static and  anisotropic content in $f(T)$ theory and obtained new spherically symmetric solutions considering a constant torsion and some particular conditions for the pressure. In this paper, still in the framework of $f(T)$ theory, new spherically symmetric solutions are obtained,  first considering the general case of an isotropic fluid and later the anisotropic content case in which the generalized conditions for the matter content are considered such that the energy density, the radial and tangential pressures depend on the algebraic $f(T)$ and its derivative $f_{T}(T)$. Moreover, we obtain the algebraic function $f(T)$ through the reconstruction method for two cases and also study a polytropic model for the stellar structure.

\end{abstract}
Pacs numbers: 04.50. Kd, 04.70.Bw, 04.20. Jb


\section{Introduction}
Through the proposal of Einstein for constructing a new version for the General Relativity (GR) \cite{case}, the Teleparallel Theory (TT) took place but has been abandoned later for many years. However, from the considerations made by Moller, the proposal for an analogy between  TT and GR has been undertaken and developed once again \cite{moller}. The GR is a theory that describes the gravitational interaction through the curvature of spacetimes. Since the Cartan manifold may possess curvature and torsion,  we can define a connection free of torsion, called Levi-Civita's connection, for which  the interaction between the gravitation and matter is described by the GR. With the progress of the TT, it can be observed that the GR, in other word, the interaction of the curvature free of the torsion with the matter, can analogously be view as a theory that possesses uniquely torsion and free of the curvature, and whose the  Riemann tensor without torsion is identically null. Then, the  TT is described  as a geometrical theory for which the Weitzenbock connection \cite{weitzenbock} generates  a torsion, which in turn defines  the action which remains invariant under the local Lorentz transformation \cite{pereira}.  From the global formulation of the TT for gravitation,  one can compare it with experiments of quantum effects in gravitational interaction of neutron interferometer \cite{pereira2,cow}.\par
With the progress of the measurements about the evolution of the universe, as the expansion and the acceleration, the dark matter and dark energy, various proposals for modifying the  GR are being tested. As the unifications theories, for the scales of low energies, it appears in the effective actions besides the Ricci scalar, the terms  $R^2$, $R^{\mu\nu}R_{\mu\nu}$ and $R^{\mu\nu\alpha\beta}R_{\mu\nu\alpha\beta}$ and the proposal of modified gravity that agrees with the cosmological and astrophysical data is $f(R)$ theory \cite{odintsov,capozziello}. The main problem that one faces with this theory is that the equation of motion is of order 4, being more complicated than the GR for any analyse. Since the GR possesses the TT  as analogous,  it has been thought the called $f(T)$  theory, $T$ being the torsion scalar, which would be the analogous to the generalizes GR, the $f(R)$ theory. The $f(T)$ theory is the generalization of the TT as we shall see later. Note also that the $f(T)$ theory is free of the curvature, i.e., it is defined  from the  Weitzenbock connection. However, it has been shown recently that this theory breaks the invariance of the local Lorentz transformations \cite{barrow}. Other recent problem is that the $f(T)$ theory appears to be dependent on the used frame, i.e., it is not covariant \cite{barrow, yapiskan}. \par
In cosmology, the $f(T)$ theory has originally been used as the source driving the inflation \cite{fiorini}. Posteriorly, it has been used as an alternative proposal for the acceleration of the universe, without requiring the introduction of the dark energy \cite{ferraro,ratbay,eric}.
 Recently, the cosmological perturbations of this theory have been analysed \cite{dutta2}. In gravitation, this theory started by obtaining  solutions of black hole BTZ \cite{fiorini2}. Note also that spherically symmetric solutions  has been obtained freshly for stars \cite{wang,boehmer}. \par
 In this paper, as well as in the simplification methods for the  equations of motion used in the GR, for obtaining the  solutions with anisotropic symmetry \cite{florides}, we propose to analyse the possibility of getting new gravitational solutions in $f(T)$ theory. Fixing the spherical symmetry and the staticity  of the metric for a matter content described by a energy momentum tensor for an anisotropic fluid, we can  impose some conditions on the functions which define the metric or on the radial pressure. Thus, one can obtain equations for the torsion scalar, and this results in the complete  determination of the Weitzenbock geometry through the differential equation that defines the torsion scalar.\par
 This paper is organized as follows. In the Section $2$, we will present a brief revision of the fundamental concepts of the Weitzenbok geometry, the action of the $f(T)$ theory and the equations of motion. In the Section $3$, we will fix the symmetries of the geometry and present the equations of the energy density, the radial and tangential pressures. The Section $4$ is devoted for obtaining new solutions in the $f(T)$ theory with constant torsion, and the matter content depending on the function $f(T)$ and its derivative. In the Section $5$, we present a summary of the reconstruction method for the static case of the theory $f(T)$, showing its effectiveness for two illustrative examples. In Section $6$ we study the stellar structure in our conjecture for the $f(T)$ theory. The conclusion and perspectives are presented in the Section $7$.
       

\section{\large The field equations from $f(T)$ theory}

As the $f(T)$ theory has been extensively studied this last months, several works have established with great consistency their mathematical bases and physical basic important concepts  for their understandability \cite{pereira,pereira3}. Let us define the notation of the Latin subscript as those related to the tetrad fields, and  the Greek one related to the spacetimes coordinates. For a general spacetimes metric, we can define the line element as
\begin{equation}
dS^{2}=g_{\mu\nu}dx^{\mu}dx^{\nu}\; .
\end{equation} 
One can describe the projection of this line element in the tangent space to the spacetimes through the matrix called tetrad as follows
\begin{eqnarray}
dS^{2} &=&g_{\mu\nu}dx^{\mu}dx^{\nu}=\eta_{ij}\theta^{i}\theta^{j}\label{1}\; ,\\
dx^{\mu}& =&e_{i}^{\;\;\mu}\theta^{i}\; , \; \theta^{i}=e^{i}_{\;\;\mu}dx^{\mu}\label{2}\; ,
\end{eqnarray} 
where $\eta_{ij}=diag[1,-1,-1,-1]$ and $e_{i}^{\;\;\mu}e^{i}_{\;\;\nu}=\delta^{\mu}_{\nu}$ or  $e_{i}^{\;\;\mu}e^{j}_{\;\;\mu}=\delta^{j}_{i}$. The square root of the metric determinant is given by  $\sqrt{-g}=\det{\left[e^{i}_{\;\;\mu}\right]}=e$.  For describing the spacetimes in terms of the tetrad matrix, we choose the connection such that the Riemann tensor vanishes identically and the  Weitzenbock connection is given by
\begin{eqnarray}
\Gamma^{\alpha}_{\mu\nu}=e_{i}^{\;\;\alpha}\partial_{\nu}e^{i}_{\;\;\mu}=-e^{i}_{\;\;\mu}\partial_{\nu}e_{i}^{\;\;\alpha}\label{co}\; .
\end{eqnarray}
For this spacetimes, the curvature is always null, while the pressure can vanish. Due to the fact that the antisymmetric part of the connection does not vanish, we can define directly the components of the connection, the tensors, the torsion and contorsion, whose components are given by
\begin{eqnarray}
T^{\alpha}_{\;\;\mu\nu}&=&\Gamma^{\alpha}_{\nu\mu}-\Gamma^{\alpha}_{\mu\nu}=e_{i}^{\;\;\alpha}\left(\partial_{\mu} e^{i}_{\;\;\nu}-\partial_{\nu} e^{i}_{\;\;\mu}\right)\label{tor}\;,\\
K^{\mu\nu}_{\;\;\;\;\alpha}&=&-\frac{1}{2}\left(T^{\mu\nu}_{\;\;\;\;\alpha}-T^{\nu\mu}_{\;\;\;\;\alpha}-T_{\alpha}^{\;\;\mu\nu}\right)\label{cont}\; ,
\end{eqnarray}
 which allow us to define new components of the tensor $S_{\alpha}^{\;\;\mu\nu}$ as
\begin{eqnarray}
S_{\alpha}^{\;\;\mu\nu}=\frac{1}{2}\left( K_{\;\;\;\;\alpha}^{\mu\nu}+\delta^{\mu}_{\alpha}T^{\beta\nu}_{\;\;\;\;\beta}-\delta^{\nu}_{\alpha}T^{\beta\mu}_{\;\;\;\;\beta}\right)\label{s}\;.
\end{eqnarray}
We are now able to define easily  the scalar that makes up the action of  $f(T)$ theory, the torsion scalar $T$.  Through  (\ref{tor})-(\ref{s}), we define the scalar torsion scalar as  
\begin{eqnarray}
T=T^{\alpha}_{\;\;\mu\nu}S^{\;\;\mu\nu}_{\alpha}\label{tore}\; .
\end{eqnarray}
The GR couples matter content with the Einstein-Hilbert action (linear function of $R$), and  the equation of motion is obtained by the minimum action principle. The $f(R)$ theory in the same way seeks the coupling with the matter content, defining instead of a linear term of the curvature scalar for the geometrical part, an arbitrary function of $R$, $f(R)$. In this paper, we will make the same considerations for coupling the geometrical part of the theory, which is the generalization of the TT, through a function depending on the torsion scalar, $f(T)$, with the Lagrangian   density of the matter content. Thus, we define the action of the $f(T)$ theory as
\begin{eqnarray}
S[e^{i}_{\mu},\Phi_{A}]=\int\; d^{4}x\;e\left[\frac{1}{16\pi}f(T)+\mathcal{L}_{Matter}\left(\Phi_{A}\right)\right]\label{action}\; ,
\end{eqnarray}
where we used the units in which $G=c=1$ and the $\Phi_{A}$ are the matter fields. Considering the action (\ref{action}) as a functional of the fields  $e^{i}_{\mu}$ and $\Phi_{A}$,  and vanishing the variation of the functional  with respect to the field  $e^{i}_{\nu}$, one obtains the following equation of motion  \cite{barrow}
\begin{eqnarray}
S^{\;\;\nu\rho}_{\mu}\partial_{\rho}Tf_{TT}+\left[e^{-1}e^{i}_{\mu}\partial_{\rho}\left(ee^{\;\;\alpha}_{i}S^{\;\;\nu\rho}_{\alpha}\right)+T^{\alpha}_{\;\;\lambda\mu}S^{\;\;\nu\lambda}_{\alpha}\right]f_{T}+\frac{1}{4}\delta^{\nu}_{\mu}f=4\pi\mathcal{T}^{\nu}_{\mu}\label{em}\; ,
\end{eqnarray}
where $\mathcal{T}^{\nu}_{\mu}$ is the energy momentum tensor, $f_{T}=\frac{\partial f(T)}{\partial T}$ and $f_{TT}=\frac{\partial^{2} f(T)}{\partial T^{2}}$. If we consider again $f(T)=a_{1}T+a_{0}$, the TT is recovered with a cosmological constant. For obtaining more general solutions in this spacetimes, we will undertake the matter content described by a energy-momentum tensor of an anisotropic fluid
\begin{eqnarray}
\mathcal{T}^{\,\nu}_{\mu}=\left(\rho+p_t\right)u_{\mu}u^{\nu}-p_t \delta^{\nu}_{\mu}+\left(p_r-p_t\right)v_{\mu}v^{\nu}\label{tme}\; ,
\end{eqnarray}
where $u^{\mu}$ is the four-velocity, $v^{\mu}$ the unit space-like vector in the radial direction, $\rho$ the energy density, $p_r$ the pressure in the direction of $v^{\mu}$ (radial pressure) and $p_t$  the pressure orthogonal to $v_\mu$ (tangential pressure). Since we are assuming an anisotropic spherically symmetric matter, on has $p_r \neq p_t$, such that their equality corresponds to an isotropic fluid sphere.\par
In the next section, we will make some considerations for the manifold symmetries in order to obtain simplifications in the equation of motion and the specific solutions of these symmetries.

\section{\large   Spherically Symmetric geometry}

The line element of a spherically symmetric and static spacetimes can be described, without loss of generality, as

\begin{equation}
dS^{2}=e^{a(r)}dt^{2}-e^{b(r)}dr^{2}-r^{2}\left(d\theta^{2}+\sin^{2}\left(\theta\right)d\phi^{2}\right)\label{ele}\; .
\end{equation} 
In order to re-write the line element (\ref{ele}) into the invariant form under the Lorentz transformations as in  (\ref{1}), we define the tetrad matrix (\ref{2}) as
\begin{eqnarray}
\left\{e^{i}_{\;\;\mu}\right\}= diag \left\{e^{a(r)/2},e^{b(r)/2},r,r\sin\left(\theta\right)\right\}\label{tetra}\; .
\end{eqnarray}
This choice of tetrad matrices is not unique, because the aim of letting the line element invariant under local Lorentz transformations, resulting in the form (\ref{1}). Other choices have been performed with non-diagonal matrices, as in references \cite{boehmer} and
 \cite{gamal}. Using  (\ref{tetra}), one can obtain $e=\det{\left[e^{i}_{\;\;\mu}\right]}=e^{(a+b)/2}r^2 \sin\left(\theta\right)$, and with (\ref{co})-(\ref{tore}), we determine the  torsion scalar and its derivatives in terms of  $r$
\begin{eqnarray}
T(r) &=& \frac{2e^{-b}}{r}\left(a^{\prime}+\frac{1}{r}\right)\label{te}\; ,\\
T^{\prime} (r)&=& \frac{2e^{-b}}{r}\left(a^{\prime\prime}-\frac{1}{r^2}\right)-T\left(b^{\prime}+\frac{1}{r}\right)\label{dte}\; ,
\end{eqnarray} 
where the prime  ($^{\prime}$) denote the derivative with respect to  the radial coordinate $r$. One can now re-write the equations of motion (\ref{em}) for an anisotropic fluid as 
\begin{eqnarray}
4\pi\rho &=& \frac{f}{4}-\left( T-\frac{1}{r^2}-\frac{e^{-b}}{r}(a'+b')\right)\frac{f_T}{2}\,,\label{dens} \\
4\pi p_{r} &=& \left(T-\frac{1}{r^2}\right)\frac{f_T}{2}-\frac{f}{4}\label{presr}\;, \\
4\pi p_{t} &=& \left[\frac{T}{2}+e^{-b}\left(\frac{a''}{2}+\left(\frac{a'}{4}+\frac{1}{2r}\right) (a'-b')\right)\right]\frac{f_T}{2}-\frac{f}{4}\,, \label{prest}\\
\frac{\cot\theta}{2r^2}T^{\prime}f_{TT}&=&0\label{impos}\;,
\end{eqnarray} 
where $p_{r}$ and $p_{t}$  are the radial and tangential pressures respectively. In the Eqs. (\ref{dens})-(\ref{prest}), we used the imposition  (\ref{impos}), which arises from the non-diagonal components $\theta\,$-$\,r$ ($2$-$1$) of the equation of motion (\ref{em}). This imposition does not appear in the static case of the GR, but making his use in (\ref{impos}), we get only the following possible solutions 
\begin{eqnarray}
T^{\prime}&=&0\Rightarrow T=T_0\label{impt}\;,\\
f_{TT}&=&0\Rightarrow f(T)=a_{0}+a_{1}T\label{impf}\;,\\
T^{\prime}&=&0,f_{TT}=0\Rightarrow T=T_0,f(T)=f(T_{0})\label{imptf}\;,
\end{eqnarray}
which always relapses into the particular case of Teleparallel Theory, with $f(T)$ a constant or a linear function. In the next section, we will determine new solutions for the $f(T)$ theory making some consideration about the matter components $\rho (r), p_{r} (r)$ and $p_{t} (r)$.



\section{New solutions for an anisotropic fluid in the Weitzenbock spacetimes}


Several works have been done in cosmology, modeling and solving some problems, using  the f(T) theory as basis.  Actually, in local and astrophysical phenomena, their is still slowly moving to obtain new solutions. Recently, Deliduman and Yapiskan \cite{yapiskan} shown that it could not exist relativistic stars, such as that of neutrons and others, in $f (T)$ theory  in $4$D, except in the linear trivial case, the usual Teleparallel Theory. However, Boehmer et al \cite{boehmer} showed that for cases where $T=0$ and $ T^{\prime} = 0 $, there exists solutions of relativistic stars. Wang \cite{wang} also shown the existence of a class of solutions, coupled with Maxwell field, in the $f(T)$ theory. Similarly, Vasquez et al \cite{vasquez} show some classes of solutions with rotation of the $f(T)$ theory in $3$D, some of them coupled with the Maxwell field. In a previously paper, we also draw the same idea and shown some classes of new solutions in $f(T)$ theory with some specific conditions for the torsion and radial pressure for anisotropic fluids \cite{stephane}. In order to extend this same idea, we will show in this paper some classes of spherically symmetric static solutions  coming from $f(T)$ theory, generalizing the condition of the matter content as depending on the algebraic functions $f(T)$ and $f_{T}(T)$ and algebraic functions of the radial coordinate $r$. Meanwhile, in order to start with the simplest cases, and later performing the generalization, we will show, first, the general case of isotropy in the next subsection.

\subsection{The isotropic case}
Before discussing the general anisotropic case, we will establish the general condition of isotropy of the solutions  the $f(T)$ theory with the metric (\ref{ele}). Taking the equality $p_{r}(r)=p_{t}(r)$, with the expressions (\ref{presr}) and (\ref{prest}), we get
\begin{equation}
T(r)=\frac{2}{r^2}+e^{-b}\left[ a^{\prime\prime}+\left(\frac{a^{\prime}}{2}+\frac{1}{r}\right)\left(a^{\prime}-b^{\prime}\right)\right]\label{tiso}.
\end{equation} 

Tanking $a^{\prime}$ and $a^{\prime\prime}$ in terms of  $T(r)$, $b(r)$ and their derivatives, through (\ref{te}), we obtain
\begin{equation}
a^{\prime}=\frac{r}{2}T(r)e^{b}-\frac{1}{r}\; ,\; a^{\prime\prime}=\frac{e^{b}}{2}\left(T+rT^{\prime}+rTb^{\prime}\right)+\frac{1}{r^2}\label{a1}\;, 
\end{equation}
which, substituted in (\ref{tiso}) yield
\begin{eqnarray}
\frac{B^{\prime}(r)}{2r}\left[1-\frac{r^2T(r)}{2B(r)}\right]+\frac{B(r)}{2r^2}\left[1+\frac{r^4T^2(r)}{4B^2(r)}\right]+\frac{2}{r^2}\left[1-\frac{r^2}{4}T(r)+\frac{r^3}{4}T^{\prime}(r)\right]=0\;,\label{isog}
\end{eqnarray}
where $B(r)=e^{-b(r)}$. This is the general equation for a model with isotropic matter content in $f(T)$ theory, for the metric (\ref{ele}), firstly obtained in this work. As the differential equation (\ref{isog}) is nonlinear and very complicated for a direct integration, with some manipulation, we have three simplified cases:
\begin{enumerate}
\item For $T(r)=0$, where $a(r)=\ln\left(r_{0}/r\right)$, the equation (\ref{isog}) becomes 
\begin{equation}
B^{\prime}(r)+\frac{1}{r}B(r)+\frac{4}{r}=0\label{B1}\;,
\end{equation} 
whose general solution is 
\begin{equation}
B(r)=e^{-b(r)}=\frac{c_{0}}{r}-4\label{b1}\;.
\end{equation}

The line element (\ref{ele}) becomes 
\begin{equation}
dS^{2}=\frac{r_{0}}{r}dt^2-\left(\frac{c_{0}}{r}-4\right)^{-1}dr^2-r^2d\Omega^2\label{sol1}\;,
\end{equation}
where $r_{0}> 0$ and $c_{0}$ is a real constant. With this, we have the following possibilities for the signature of the metric: a) when $c_{0}>4r$, the signature is $(+---)$; b) when $c_{0}<4r$, the signature is $(++--)$\footnote{ Metrics with two timelike coordinates  were studied in strings models with Timelike T-Duality \cite{hull},  in branes \cite{gibbons}, in Sigma model \cite{gerard} and other physical applications \cite{bars}.}. This solution was first obtained by Boehmer et al \cite{boehmer}, but with a general equation of isotropy different from ours, and with an incorrect limit\footnote{They choose the torsion as constant and make the limit $r_{0}^2T_0<<1$ in an equation similar to (\ref{isog}), which clearly does not lead to (\ref{B1}), due to the terms where there exists $B^{-1}(r)$ in (\ref{isog}).}. 
\par
Boehmer et al obtained this solution by using a wrong limit, and have not drawn up any analysis. This  solution could be view as a wormhole. We can observe this by following the same process as in \cite{stephane}. This line element (\ref{ele}) can be put in the form  
\begin{equation}
dS^{2}=e^{a(r)}dt^{2}-dl^{2}-r^{2}(l)d\Omega^{2}\label{elw}\;,
\end{equation}
where $a(r)$ is denoted redshift function, and through the redefinition $\beta (r)=r\left[1-e^{-b(r)}\right]$, with $b(r)$ being the metric function given in (\ref{ele}), $\beta(r)$ is called  shape function. Therefore, the conditions of existence of a traversable wormhole are: a) the function $r(l)$ must possess a minimum value $r_{1}$ for $r$, which imposes ${d^2r}(l)/dl^{2}>0$; b) $\beta(r_{1})=r_{1}$;  c) $a(r_{1})$ has a finite value; and finally d) $d\beta(r)/dr|_{r=r_{1}}\leqslant  1$.
\par
The solution (\ref{sol1}) satisfies the conditions  a), for  $c_{0}<0$ (signature $(++--)$), b) and  c), but not the condition d), because $\beta^{\prime}(r_{1})=5>1$,  and then seems to be a wormhole, but not a traversable one. However, as the signature must be  $(++--)$, it could not be an usual wormhole, with the evolution given by Einstein-Rosen bridge, and then is discarded because of being non-physical. Now, the solution with $c_{0}>4r$ (signature $(+---)$), is not a wormhole, but a possible solution for a null torsion scalar.
\par
The energy density and the pressure (isotropic case) are given by
\begin{eqnarray}
\rho (r)&=&\frac{f(0)}{16\pi}+\frac{5f_{T}(0)}{8\pi r^2}\label{dens1}\;,\\
p_{r}(r)&=&p_{t}(r)=-\frac{f(0)}{16\pi}-\frac{f_{T}(0)}{8\pi r^2}\label{pres1}\;.
\end{eqnarray}  
This solution presents a matter content divergent in $r= 0$, then is called singular. The matter content must satisfy the weak energy condition (WEC) and
the null energy condition (NEC), given by $\rho (r)\geqslant 0$, $\rho (r)+p_{r}(r)\geqslant 0$ and  $\rho (r)+p_{t}(r)\geqslant 0$. Then, for guaranteeing    the WEC and NEC, we need to have $f(0),f_{T}(0)\geqslant 0$, or  $f(0)\leqslant 0$ and $f_{T}(0)\geqslant 0$, with $0\leqslant r\leqslant r_{+}$, where $r_{+}=\sqrt{-10f_{T}(0)/f(0)}$. 
\item Choosing the condition 
\begin{equation}
a^{\prime}(r)=-\frac{1}{r}+c_{0}\;,\;c_{0}\in\mathcal{R}\label{cond1}\;,
\end{equation}
that generalizes the previous case, where we had $T=0$, through the equation (\ref{te}), (\ref{isog}) becomes
\begin{eqnarray}
B^{\prime}(r)(1+c_{0}r)+B(r)\left(\frac{1}{r}+c^{2}_{0}r-4c_{0}\right)+\frac{4}{r}=0\label{B2}\;.
\end{eqnarray}

The general solution of this equation is given by  
\begin{eqnarray}
B(r)&=& r^{-1}e^{-c_{0}r}\left(1+c_{0}r\right)^{6}\left(c_{1}-\frac{k(r)}{180ec_{0}}\right)+\frac{1}{180c_{0}r}\Big( 154+51c_{0}r+28c_{0}^2r^2+16c_{0}^3r^3+\nonumber\\
&&+6c_{0}^4r^4+c_{0}^5r^5\Big)\label{b2}\;,
\end{eqnarray}
where $k(r)=exp\left(-\int^{\infty}_{-r}z^{-1}e^{-z}dz \right)$ and $c_{1}$ is a real constant. The solution of $a(r)$, coming from (\ref{cond1}), is given by 
\begin{equation}
e^{a(r)}=\frac{r_{0}}{r}e^{c_{0}r}\label{a2}\;.
\end{equation}
The expression of energy density, the radial and tangential pressures are too long and can not be written here. However, it is important to note that they are singular in $r= 0$. This is a new isotropic solution obtained for the first time in this work.

\item Taking the so-called quasi-global coordinate condition
\begin{equation}
a(r)=-b(r)\label{cond2}\;,
\end{equation}
from (\ref{dens}) and (\ref{presr}) one obtains the equality
\begin{equation}
p_{r}(r)=-\rho (r)\label{eq1}\;.
\end{equation}
The isotropy requires $p_{r}(r)=p_{t}(r)$, which, from (\ref{presr}) and (\ref{prest}) yields (\ref{tiso}). Replacing the condition (\ref{cond2}) in (\ref{te}) and equating with the expression (\ref{tiso}), we get the following differential equation
\begin{eqnarray}
b^{\prime\prime}-\left(b^{\prime}\right)^{2}+\frac{2}{r^2}\left(1-e^{b}\right)=0\label{biso}\;.
\end{eqnarray}

The solution of this equation is
\begin{equation}
e^{a(r)}=e^{-b(r)}=1-\frac{c_{0}}{r}+\frac{c_{1}}{3}r^2\label{desol}\;,
\end{equation}
where $c_{0},c_{1}\in\mathcal{R}$. This solution (\ref{desol}) behaves as the equation of state of dark energy, $p_{r}(r)=p_{t}(r)=-\rho (r)$, with
\begin{eqnarray}
\rho (r)=\frac{a_{0}-2a_{1}c_{1}}{16\pi}\label{dedens}\;,
\end{eqnarray}
where we took into account the imposition  (\ref{impf}). This is a new black hole solution obtained in this work. This solution is similar to the  S-(A)dS one, for $c_{0}=2M$ and $c_{1}=-\Lambda$. The conditions $\rho (r)+p_{r}(r)\geqslant 0$ and $\rho (r)+p_{t}(r)\geqslant 0$ are always satisfied, since $\rho (r)+p_{r}(r)=\rho (r)+p_{t}(r)=0$ in this case. The condition $\rho\geqslant 0$ impose  $a_{0}\geqslant 2a_{1}c_{1}$, for $a_{0}>0$, and $2|a_{1}c_{1}|\geqslant |a_{0}|$, for $a_{0}<0$ and $sign\left(a_{1}c_{1}\right)=-1$. Here, the torsion scalar (\ref{te}) cannot be a constant. 
\par
A similar solution to this one was obtained in our previous paper \cite{stephane}. It was obtained as a particular case of anisotropic solution for the choice of the constant radial pressure, which, by the quasi-global coordinate condition, resulted in a isotropized solution. But here, we have obtained a general isotropic solution for the quasi-global condition (\ref{cond2}), which leads to a matter content that satisfies the equation of state of the dark energy, with a constant energy density (\ref{dedens}). Comparing this later with our particular case previously obtained (solution ($55$) in \cite{stephane}), it appears that the constant pressure $p_{r}$ must be equal to the energy density (\ref{dedens}), which fixes the pressure in terms of the constants of the algebraic function $f(T)$ and $c_{1}$.

\end{enumerate}

\subsection{The anisotropic content case}
Now, as already shown in our previous paper \cite{stephane}, the equations of motion (\ref{dens})-(\ref{prest}) allow us to establish the following generalized conditions
\begin{eqnarray}
\rho (r)&=&\frac{g_{1}(r)}{16\pi}f(T)+\frac{g_{2}(r)}{8\pi}f_{T}(T)+\frac{g_{3}(r)}{16\pi}\label{gcond1}\, ,\\
p_{r}(r)&=& \frac{g_{4}(r)}{16\pi}f(T)+\frac{g_{5}(r)}{8\pi}f_{T}(T)+\frac{g_{6}(r)}{16\pi}\label{gcond2}\, ,\\
p_{t}(r)&=&\frac{g_{7}(r)}{16\pi}f(T)+\frac{g_{8}(r)}{8\pi}f_{T}(T)+\frac{g_{9}(r)}{16\pi}\label{gcond3}\, ,
\end{eqnarray}
where the functions $g_{i}(r)$, with $i=1,...,9$, are algebraic functions of only the radial coordinate $r$. These generalized conditions for the matter content are first used here, resulting in new solutions originally obtained in this work. We will distinguish three main cases here:
\begin{enumerate}


\item When the energy density obeys the condition (\ref{gcond1}), equating (\ref{dens}) and (\ref{gcond1}),  and taking into account the imposition (\ref{impf}), we get
\begin{eqnarray}
T(r)=\frac{a_{0}}{a_{1}}\left[2k_{1}(r)-1\right]-2k_{1}(r)\left[g_{2}(r)+\frac{g_{3}}{2a_{1}}-\frac{1}{r^2}-\frac{e^{-b}}{r}\left(a^{\prime}+b^{\prime}\right)\right]\label{t3}\; ,
\end{eqnarray}
where $k_{1}^{-1}(r)=g_{1}(r)+1$. Three important sub-cases can be observed:
\begin{enumerate}

\item Making use of the coordinate condition (\ref{cond1}), we obtain directly the solution (\ref{a2}) for $a(r)$. Equating (\ref{te}) and (\ref{t3}), then substituting (\ref{a2}), we get the following differential equation
\begin{eqnarray}
\left(e^{-b}\right)^{\prime}&+&x_{1}(r)\left(e^{-b}\right)+y_{1}(r)=0\label{B3}\;,\\
x_{1}(r)&=&c_{0}\left(\frac{1}{k_{1}(r)}-1\right)+\frac{1}{r}\label{x1}\;,\\
y_{1}(r)&=&r\left[g_{2}(r)+\frac{g_{3}(r)}{2a_{1}}-\frac{1}{r^2}-\frac{a_{0}}{a_{1}}\left(1-\frac{1}{2k_{1}(r)}\right)\right]\; ,\label{y1}  
\end{eqnarray}
whose general solution is 
\begin{eqnarray}
e^{-b(r)}&=& \exp\left(-\int x_{1}(r)dr\right)\left[c_{1}-\int\exp\left(\int x_{1}(r)dr\right)y_{1}(r)dr\right]\;\;,\label{b3}
\end{eqnarray}
where $c_{1}\in\mathcal{R}$, $x_{1}(r)$ and $y_{1}(r)$ are given in (\ref{x1}) and (\ref{y1}) respectively. 
\par
A particular case is when we choose  
\begin{eqnarray}
k_{1}(r)=\frac{c_{0}r}{c_{0}r-1}\;,\;g_{2}(r)=\displaystyle\sum_{n}h_{(n)}r^{n-1}-\frac{g_{3}(r)}{2a_{1}}+\frac{1}{r^{2}}+\frac{a_{0}}{a_{1}}\left(\frac{c_{0}r+1}{2c_{0}r}\right)\;.
\end{eqnarray} 
In this case, we get $x_{1}(r)=0$ and $y_{1}(r)=\displaystyle\sum_{n}h_{(n)}r^{n}$, from which, using (\ref{b3}), yields 
\begin{equation}
e^{-b(r)}=-h_{(-1)}\ln r-\displaystyle\sum_{n}\frac{h_{(n)}}{(n+1)}r^{n+1}\;, n\neq -1\;.\label{b3.1}
\end{equation}
Now, if we use only the terms in which $n=h_{(-1)}=0$ and $n=-2$, we regain a wormhole already obtained in \cite{stephane}, for $h_{(0)}=-a_{0}/2a_{1}c_{0}$ and $h_{(-2)}=1/c_{0}$.\par
 
Another particular case would be a generalization of several classes of traversable wormholes that connect two non-asymptotically flat regions. We explicit here two cases:
\begin{enumerate}
\item When the unique terms of (\ref{b3.1}) are for the orders  $r$ and  $r^2$ ($h_{(-1)}=0$), we get
\begin{equation}
e^{-b(r)}=-h_{(0)}r-\frac{h_{(1)}}{2}r^2\label{b3-2}\; .
\end{equation}
This solution is a traversable wormhole. We can show this as follows: The shape function and its derivative are given by 
\begin{equation}
\beta (r)=r\left[1+r\left(h_{(0)}+\frac{h_{(1)}}{2}r\right)\right]\; ,\; \beta^{\prime}(r)=1+r\left(2h_{(0)}+\frac{3h_{(1)}}{2}r\right)\;.\label{beta1}
\end{equation}
From (\ref{b3-2}), we take  $dr/dl=\sqrt{e^{-b(r)}}=0$,
 and solving, we obtain $r_{1}=-2h_{(0)}/h_{(1)}$. As $d^2r/dl^2=[\beta(r)-r\beta^{\prime}(r)]/2r^2>0$ \cite{matt} is a condition for the minimum and in this case we get $d^2r/dl^2|_{r_{1}}=h_{(0)}/2$, which is greater than zero for $h_{(0)}>0$, being  $r_{1}$ a minimum for $r(l)$. The redshift function $a(r)$ has a finite value in  $r_{1}$. The shape function satisfies  $\beta(r_{1})=r_{1}$ and  $\beta^{\prime}(r_{1})=1+(2h_{(0)}^2/h_{(1)})< 1$, for  $h_{(1)}<0$. In general, in order to get consecutive orders  $r^p$ and $r^{p+1}$, in (\ref{b3.1}), we obtain traversable wormholes with minimum in  $r_{1}=-(p+2)h_{(p)}/(p+1)h_{(p+1)}$, with $h_{(p)}>0$ and $h_{(p+1)}<0$. 


\item When the have only the terms of orders $r$ and $r^3$ in  (\ref{b3.1}), we get 
\begin{equation}
e^{-b(r)}=-h_{(0)}r-\frac{h_{(2)}}{3}r^3\label{b3.3}\; .
\end{equation}
This solution is a traversable wormhole. The function $r(l)$ has a minimum in  $r_{1}=\sqrt{-3h_{(0)}/h_{2}}$, for $h_{(0)}>0$. The redshift function $a(r)$ has a finite value in  $r_{1}$. The shape function satisfies  $\beta(r_{1})=r_{1}$ and  $\beta^{\prime}(r_{1})=1+(6h_{(0)}^2/h_{(2)})< 1$, for $h_{(2)}<0$. In general, for obtaining consecutive orders  $r^p$ and  $r^{p+2}$, in (\ref{b3.1}), we get traversable wormholes with minimum in  $r_{1}=\sqrt{-(p+3)h_{(p)}/(p+1)h_{(p+2)}}$, with $h_{(p)}>0$ and $h_{(p+2)}<0$. We also have a multitude of other solutions of traversable wormholes in (\ref{b3}), that connect two non-asymptotically flat regions.
 
\end{enumerate} 

\item When the condition of coordinates is given by the  quasi-global coordinate (\ref{cond2}), the equation (\ref{te}) yields the following equation
\begin{equation}
\left(e^{-b}\right)^{\prime}+\frac{1}{r}\left(e^{-b}\right)-\frac{r}{2}T(r)=0\label{bt}\;,
\end{equation}
whose general solution is 
\begin{equation}
e^{a(r)}=e^{-b(r)}=\frac{c_{0}}{r}+\frac{1}{2r}\int r^2 T(r)dr\;,\;c_{0}\in\mathcal{R},\label{b3.2}
\end{equation}
which, from (\ref{t3}), the torsion scalar is given by 
\begin{eqnarray}
T(r)=\frac{a_{0}}{a_{1}}\left[2k_{1}(r)-1\right]-2k_{1}(r)\left[g_{2}(r)+\frac{g_{3}}{2a_{1}}-\frac{1}{r^2}\right]\label{t3.2}\; .
\end{eqnarray}
We can show that a particular case that highlights the generalization of this solution. Taking
\begin{equation}
k_{1}(r)=\frac{\frac{a_{0}}{a_{1}}+\displaystyle\sum_{n} h_{(n)}r^{n}}{2\left[\frac{a_{0}}{a_{1}}-g_{2}(r)-\frac{g_{3}(r)}{2a_{1}}+\frac{1}{r^{2}}\right]}\label{k1}\;,
\end{equation}
where $h_{(n)}$ are real constants  and  $n\in\mathbb{Z}$, the torsion scalar in (\ref{t3.2}) becomes  $T(r)=\displaystyle\sum_{n} h_{(n)}r^{n}$. Substituting it in (\ref{b3.2}), we get the following particular case
\begin{equation}
e^{a(r)}=e^{-b(r)}=\frac{c_{0}}{r}+\frac{h_{(-3)}}{2r}\ln r +\frac{1}{2}\displaystyle\sum_{n} \frac{h_{(n)}}{(n+3)}r^{n+2}\label{b3.3}\;,
\end{equation}
with $n\neq -3$. Two specific cases of this particular case are: a) when $h_{(-3)}=n=0$, we regain the same  result as obtained in \cite{stephane}, with $h_{(0)}=T_{0}$; b) when we only have the terms for the values of  $n=h_{(-3)}=0$, $n=-2$ and $n=-4$. In this case the solution is of type  Reissner-Nordstrom-(Anti) de Sitter (RN-(A)dS), with the mass  $M=-c_{0}/2$, the electric charge $q^2=-h_{(-4)}/2$ ($h_{(-4)}<0$) and the cosmological constant $\Lambda =-h_{(0)}/2$. 
\par
This particular case reproduce various known terms, with respect to the power of the radial coordinate $r$, as the linear term \cite{kazanas}, the logarithmic term (in GR \cite{kuchowicz}, in $f(R)$ theory \cite{mariafelicia} and in other modified gravities \cite{stephane2}), the term of fourth power \cite{tolman,mehra}, in that of the nth order \cite{tolman,wyman,kuchowicz} among others. In fact, the general case is more comprehensive.   
\par
If we take the particular case of the example b), with the unique terms $h_{(-2)}$ and $h_{(0)}$, for $h_{(-2)}=2$ (type S-dS), we obtain the line element as 
\begin{equation}
dS^2=\left(1+\frac{c_0}{r}+\frac{h_0}{6}r^2\right)dt^2-\left(1+\frac{c_0}{r}+\frac{h_0}{6}r^2\right)^{-1}dr^2-r^2d\Omega^2\label{ex}\;.
\end{equation}    
The horizon is obtained through  $g_{00}(r_H)=0$. The energy density in this case is given by  $\rho=(a_0-a_1h_0)/16\pi$. Then, the total mass is given by  $M=4\pi\int^{r_H}_{0}\rho r^2dr=(4\pi/3)\rho r_H^3$. 
Making the match of the interior metric with an exterior one, of type S-dS, where $c_0=-2M$, the horizon is then obtained on the form  $r_H=\sqrt{6/[a_0-h_0 (1+a_1)]}$. With $g_{00}(r_H)=0$, we write the total  mass and its differential as 
\begin{eqnarray}
M=\frac{r_H}{2}\left(1+\frac{h_0}{6}r_H^2\right)\label{m1}\;,\\
dM=\frac{dr_H}{2}\left(1+\frac{h_0}{2}r_H^2\right)\;.\label{m2}
\end{eqnarray}
The Hawking temperature, the entropy and its differential can be calculated through (\ref{ex}), as
\begin{eqnarray}
T_H&=&\frac{g_{00}^{\prime}}{4\pi\sqrt{-g_{00}g_{11}}}\Big|_{r=r_H}=\frac{1}{4\pi}\left(\frac{2M}{r_H^2}+\frac{h_0}{3}r_H\right)\;,\label{th1}\\
S&=&\frac{1}{4}A_H=\frac{1}{4}\int^{\pi}_{0}\int_{0}^{2\pi}\sqrt{g_{22}g_{33}}d\theta\phi \Big|_{r=r_H}=\pi r_H^2\;,\label{s1}\\
dS&=&2\pi r_H dr_H\;.\label{s2}
\end{eqnarray}
From the expressions (\ref{m2}), (\ref{th1}) and (\ref{s2}), taking into account  (\ref{m1}), we can show that the solution  (\ref{ex}) obeys the first law  of thermodynamics for black holes.
\begin{equation}
dM=T_HdS\;.
\end{equation}
This fact is not surprising. Recently, Miao Li and collaborators \cite{miao} have demonstrated that it can exist a violation of the first law of thermodynamics in $f(T)$ theory, when the invariance by the local Lorentz transformation is broken, due to the term $f_{TT}(T)$ in the equations of motion (\ref{em}). When the first law is violated, there exists entropy production. But for our specific case of the choice of a set of diagonal tetrads in
 (\ref{tetra}), it appears the imposition (\ref{impf}), which eliminates the terms of $f_{TT}(T)$ in the equations of motion. Then, we observe that the first law is always  obeyed, as shown by Mian Li and collaborators  \cite{miao}, with the entropy  $S=(f_{T}[T(r_H)]/4)A_H$. Here, we obtained the entropy by  (\ref{s1}), but this is due to the fact that we can use the symmetry of temporal translation in the action  (\ref{action}), and put $f(T)=a_1[T+(a_0/a_1)]$ ($a_0/a_1\rightarrow \widehat{a}_0$), then, eliminating the constant $a_1$ in $dt$. With this, we get $f_{T}[T(r_H)]=1$, conciliating our result with the general conjecture of Miao and collaborators. The same thermodynamic analysis can be made for other black holes solutions obtained in this work. We will just explain  this particular case, due to the wide rang of classes of solutions obtained in this work. 
 

\item When the coordinate condition is given by 
\begin{equation}
b^{\prime}=-\frac{1}{r}\label{cond3}\;,
\end{equation}
 the equation (\ref{te}) yields
\begin{equation}
T(r)=\frac{2}{r_{0}r}+\frac{2}{r_{0}}a^{\prime}\label{t3.3}\;,
\end{equation}
where $r_{0}$ is a positive integration constant. Substituting (\ref{t3.3}) into (\ref{t3}), we get
\begin{eqnarray}
a(r)=\int \left[\frac{1}{r}\left(\frac{k_{1}(r)+1}{k_{1}(r)-1}\right)+\frac{r_{0}k_{1}(r)}{(k_{1}(r)-1)}\left(g_{2}(r)+\frac{g_{3}(r)}{2a_{1}}-\frac{1}{r^2}\right)-\frac{r_{0}a_{0}}{2a_{1}}\left(\frac{2k_{1}(r)-1}{k_{1}(r)-1}\right)\right]dr\label{a4}\,,
\end{eqnarray}
and
\begin{equation}
e^{b(r)}=\frac{r_{0}}{r}\label{b3.4}\;.
\end{equation}

Now, for a particular , if we choose \begin{equation}
k_{1}(r)=\frac{\frac{a_{0}}{a_{1}}+\frac{2}{r_{0}}\ln\left(\displaystyle\sum_{n} h_{(n)}r^{n}\right)}{2\left[\frac{a_{0}}{a_{1}}-g_{2}(r)-\frac{g_{3}(r)}{2a_{1}}+\frac{1}{r^{2}}\right]}\label{k12}\;,
\end{equation}
we regain the results  mentioned in the previous  item according to the choice of the constants $r_{0}$ and $h_{(n)}$, but with $b(r)$ given in (\ref{b3.4}) and $a(r)$ in (\ref{b3.3}), and which reproduces the various  terms of this case.
\end{enumerate}


\item When the tangential pressure obeys the condition (\ref{gcond3}), equating (\ref{prest}) and (\ref{gcond3}), considering the imposition (\ref{impf}), and making $g_{7}(r)=k_{2}^{-1}(r)$, we obtain
\begin{eqnarray}
T(r)&=&k_{2}(r)\left[2\left(e^{-b}\left(\frac{a^{\prime\prime}}{2}+\left(\frac{a^{\prime}}{4}+\frac{1}{2r}\right)(a^{\prime}-b^{\prime})\right)-g_{8}(r)\right)-\frac{g_{9}(r)}{a_{1}}\right]-\frac{a_{0}}{a_{1}}\left[k_{2}(r)+1\right]\label{t4}\,.
\end{eqnarray} 
A direct solution can be obtained by taking the condition (\ref{cond1}) and  substituting into (\ref{t4}), that leads to
\begin{eqnarray}
&&x_{2}(r)\left(e^{-b}\right)^{\prime}+y_{2}(r)\left(e^{-b}\right)-z_{2}(r)=0\label{B4}\;,\\
&&x_{2}(r)=\frac{k_{2}(r)}{4}\left(c_{0}+\frac{1}{r}\right)\;,\;y_{2}(r)=\frac{k_{2}(r)}{4}\left(\frac{c_{0}^2}{4}+\frac{1}{2r^2}\right)-\frac{c_{0}}{r}\,,\label{x2}\\
&&z_{2}(r)=k_{2}(r)\left(g_{8}(r)+\frac{g_{9}(r)}{2a_{1}}\right)+\left(\frac{a_{0}}{2a_{1}}\right)\left[k_{2}(r)+1\right]\;.\label{z2}
\end{eqnarray}

The general solution of (\ref{B4}) is
\begin{equation}
e^{-b(r)}=exp\left(-\int \frac{y_{2}(r)}{x_{2}(r)}dr\right)\left[c_{1}+\int \left(exp\left[\int \frac{y_{2}(r)}{x_{2}(r)}dr\right]\right)\frac{z_{2}(r)}{x_{2}(r)}dr\right]\label{b4}\;,
\end{equation}
where $c_{1}\in\mathcal{R}$, $x_{2}(r),y_{2}(r)$ and $z_{2}(r)$ are given in (\ref{x2}) and (\ref{z2}). For the particular case in which 
\begin{eqnarray}
k_{2}(r)=\frac{16c_{0}r}{c_{0}^2 r^2+2}\;,\; g_{8}(r)=\frac{(c_{0}r+1)}{4}\displaystyle\sum_{n} h_{(n)}r^{n-1}-\frac{a_{0}}{2a_{1}}\left(1+\frac{c_{0}^2 r^2+2}{16c_{0}r}\right)-\frac{g_{9}(r)}{2a_{1}}\;,
\end{eqnarray}
we get 
\begin{equation}
e^{-b(r)}=h_{(-1)}\ln r +\displaystyle\sum_{n} \frac{h_{(n)}}{(n+1)}r^{n+1}\;,
\end{equation}
 with $n\neq -1$. This solution reproduces the cases of  (\ref{b3.1}), substituting $h_{p}\rightarrow -h_{p}$.

\item When the radial pressure obeys the condition (\ref{gcond2}). Equating (\ref{presr}) and (\ref{gcond2}), tanking into account the imposition (\ref{impf}), putting  $k_{3}^{-1}(r)=1-g_{4}(r)$ we get
\begin{equation}
T(r)=\frac{a_{0}}{a_{1}}\left[2k_{3}(r)-1\right]+2k_{3}(r)\left[\frac{1}{r^2}+g_{5}(r)+\frac{g_{6}(r)}{2a_{1}}\right]\label{t5}\;.
\end{equation}

Here we observe three important cases: 
\begin{enumerate}

\item For the coordinate condition (\ref{cond1}), we get
\begin{eqnarray}
e^{a(r)}=\frac{r_{0}}{r}e^{c_{0}r}\; ,\; e^{b(r)}=\frac{2c_{0}}{rT(r)}\label{b5}\;,
\end{eqnarray}
where $T(r)$ is given in (\ref{t5}).


\item For the coordinate condition (\ref{cond2}), one gets the equation (\ref{bt}), whose solution is (\ref{b3.2}), but in this case $T(r)$ is given by (\ref{t5}).


\item For the coordinate condition  (\ref{cond3}), whose solution is (\ref{b3.4}), we get the equation (\ref{t3.3}), which results into 
\begin{equation}
e^{a(r)}=\frac{r_{1}}{r}\exp\left(\frac{r_{0}}{2}\int T(r)dr\right)\label{a5}\;,
\end{equation}
where $r_{0},r_{1}>0$ and $T(r)$ is given by (\ref{t5}).\par
We could reclaim the particular cases $2$-$5$, of the radial pressure, treated previously in \cite{stephane}, making $g_{4}(r)=g_{5}(r)=g_{6}(r)=0$ ($p_{r}(r)=0$), $g_{4}(r)=[8\pi p_{r}/(a_{0}+8\pi p_{r})],g_{5}(r)+(g_{6}(r)/2a_{1})=-g_{4}(r)/r^2$ ($p_{r}(r)=p_{r}\in\mathcal{R}$), $g_{4}(r)=(16\pi/c_{0}),g_{5}(r)=g_{6}(r)=0$ ($p_{r}(r)=f(T)/c_{0}$) and $g_{4}(r)=g_{6}(r)=0,g_{5}(r)=\eta /r^2$ ($p_{r}(r)=(\eta /r^2)f_{T}(T)$), respectively. But in general, we have much more comprehensive solutions here. 
\end{enumerate}
\end{enumerate}
\section{Reconstruction in static f(T) theory}
A method widely used in cosmology is called reconstruction. This method stems from the introduction of an auxiliary field for the reconstruction of the algebraic function of the main action, as in the case of the theory $f(R)$ for example \cite{reconstruction1}.  
\par
We can briefly present this method as follows. Considering the algebraic function
\begin{equation}
f(T)=P\left(\varphi\right)T+Q\left(\varphi\right)\label{fr}\;,
\end{equation} 
the functional variation of the action  (\ref{action}), with respect to  $\varphi$, is given by 
\begin{equation}
\frac{\delta S}{\delta \varphi}=\frac{e}{16\pi}\left[\frac{dP}{d \varphi}T+\frac{d Q}{d\varphi}\right]=0\label{econst}\;.
\end{equation}  
Solving this equation, we get $\varphi\equiv\varphi(T)$, then,  $f(T)=P[\varphi(T)]T+Q[\varphi(T)]$. Hence, we have the following identities
\begin{eqnarray}
f_{T}(T)&=&P+\left(\frac{dP}{d\varphi}T+\frac{dQ}{d\varphi}\right)\frac{d\varphi}{dT}=P[\varphi(T)]\label{ftr}\;,\\
f_{TT}(T)&=&\frac{dP[\varphi(T)]}{dT}\label{fttr}\;.
\end{eqnarray} 
Having in hand the equations (\ref{fr}), (\ref{ftr}) and (\ref{fttr}), and substituting into (\ref{dens})-(\ref{impos}), we get 
\begin{eqnarray}
4\pi\rho&=&\frac{P}{2}\left[\frac{1}{r^2}+\frac{e^{-b}}{r}\left(a^{\prime}+b^{\prime}\right)-\frac{T}{2}\right]+\frac{Q}{4}\;,\label{densr}\\
4\pi p_{r}&=&\frac{P}{2}\left[\frac{T}{2}-\frac{1}{r^2}\right]-\frac{Q}{4}\label{presrr}\;,\\
4\pi p_{t}&=& \frac{P}{4}e^{-b}\left[ a^{\prime\prime}+\left(\frac{a^{\prime}}{2}+\frac{1}{r}\right)(a^{\prime}-b^{\prime})\right]-\frac{Q}{4}\label{prestr}\;,\\
\frac{dP}{dr}&=&0\label{imposr}\;.
\end{eqnarray}
The reconstruction can be performed directly, making use of the auxiliary field $\varphi=r$. The equation  (\ref{imposr}) reports that  $P\in\Re$, which is a constant. Thus, for reconstructing $f(T)$, we have to determine $Q$ in (\ref{fr}). In order to avoid repetition, we will show a simplest case in which the radial pressure is constant, and the process of re-obtaining the isotropic cases treated in the subsection $4.1$, since we do not want to re-obtain all the solutions shown in this paper. However, this method may be used for re-obtaining or reconstructing, when the inversion $r\equiv r(T)$ is possible, for the $f(T)$ theory in the static case.\par
Let us draw up to cases here:
\begin{enumerate}

\item When the radial pressure (\ref{presrr}) is a constant $p_{r}\in\Re$. In this case, from (\ref{presrr}), we get
\begin{equation}
Q=PT-\frac{2P}{r^2}-16\pi p_{r}\;,
\end{equation}
which, substituting in (\ref{fr}), yields 
\begin{equation}
f(T)=2P\left(T-\frac{1}{r^2}\right)-16\pi p_{r}\label{fr1}\;.
\end{equation}
Differentiating (\ref{fr1}) with respect to the torsion scalar  $T$, and equating to (\ref{ftr}), we obtain 
\begin{equation}
f_{T}(T)=2P+\frac{4P}{r^3}\frac{dr}{dT}=P\;,
\end{equation}
whcih, integrating, leads to 
\begin{equation}
T(r)=T_{0}+\frac{2}{r^2}\label{tr1}\;.
\end{equation}
This result agrees with that of \cite{stephane}. Now,
if we reverse this equation, for obtaining $r(T)$, one may substitute it in (\ref{fr1}), getting 
\begin{equation}
f(T)=PT+PT_{0}-16\pi p_{r}\label{fr1.1}\;.
\end{equation}
We then obtain $Q=PT_{0}-16\pi p_{r}$, and finalise  the reconstruction of the algebraic function  $f(T)$ in (\ref{fr1.1}). In this case, the solution is linear, in agreement with the constraint (\ref{impf}).

\item For the case in which 
\begin{equation}
4\pi\frac{dp_{r}}{dT}=g(T)\label{cond4}\;,
\end{equation}
where $g(T)$ is an algebraic function. Integrating  (\ref{cond4}) and equating with (\ref{presrr}), one gets
\begin{equation}
Q=PT-\frac{2P}{r^2}-4\int g(T)dT\label{q1}\;.
\end{equation}
Substituting  (\ref{q1}) into (\ref{fr}), one gets 
\begin{equation}
f(T)=2P\left(T-\frac{1}{r^2}\right)-4\int g(T)dT\label{fr2}\;,
\end{equation}
which, differentiating with respect to $T$, equating with (\ref{ftr}) and integrating, yields
\begin{equation}
r^{-2}=\frac{1}{2}(T-T_{0})-\frac{2}{P}\int g(T)dT\label{ert}\;.
\end{equation}
Inserting (\ref{ert}) into (\ref{fr2}), the algebraic function is reconstructed as 
\begin{equation}
f(T)=PT+PT_{0}\;,\label{fr2.1}
\end{equation}
as $Q=PT_{0}$ and is $f(T)$ linear, obeying again the imposition  (\ref{impf}). 
The radial pressure can be given as a power-law of the torsion scalar, for example, if  $g(T)=\displaystyle\sum_{n} h_{(n)}T^n$, and getting  $4\pi p_{r}=h_{(-1)}\ln T+\displaystyle\sum_{n} [h_{(n)}/(n+1)]T^{n+1}$, with $n\neq -1$. 
\end{enumerate}
We can obtain several cases here. Our General solutions can also be regained by this method of reconstruction. An example is when we make use of the general condition of isotropy $p_{r}=p_{t}$, matching (\ref{presrr}) with 
(\ref{prestr}),  leading to the expression (\ref{tiso}). So, we can follow the same procedures of the subsection $4.1$ and regain all previously cases treated there.\par 

The reconstruction method presented in this section for reconstructing the algebraic function $f(T)$, is the same usually used and well known in cosmology, where there is already a pre-established metric, that of FRW. Moreover, we  emphasize here that in gravitation, the situation can be view on other way, since in some cases, we also need to obtain a metric. Then, the reconstruction method appears to be of two interests. It can be used for reconstructing the algebraic function $f(T)$, the static case, or for obtaining new solutions (metric), because the equations of motion (\ref{dens})-(\ref{impos}) are more easily solved than the (\ref{densr})-(\ref{imposr}) ones, where this method is not used.\par
An observation for which we pay special attention here is that the choice of a set of diagonal tetrad for a spherically symmetric and static metric, always results in the imposition (\ref{impos}), which always results in the possibilities (\ref{impt})-(\ref{imptf}). Our results are consistent with these possibilities.

\section{Stellar structure in hydrostatic equilibrium}
In this section, we will study the stellar structure for the solutions coming from the equations of motion (\ref{dens})-(\ref{impos}). To do this, we have to take into account the results previously obtained in this paper. The first is that a choice of the diagonal tetrad, as we have done in (\ref{tetra}), leads to the equations of motion that impose a linear algebraic function $f(T)$, as in (\ref{impf}). Hence, we consider the case without cosmological constant, $a_{0}=0$, and simplify setting $a_{1}=1$, and then we have $f(T)=T$ and $f_T=1$. This does not lead us to a loss of generality in this case, since the imposition (\ref{impf}) comes from the equations of motion.\par
In the second way, we will take the line element (\ref{ele}) as
\begin{eqnarray}
dS^{2}=e^{2\Phi (r)}dt^2-\left[1-\frac{2M(r)}{r}\right]^{-1}dr^2-r^2d\Omega^2\,\label{ele1}
\end{eqnarray}
where $M(r)$ is the mass function of the star, given by the expression
\begin{equation}
M(r)=4\pi\int r^2\rho (r)dr\label{mass}
\end{equation}
and $\Phi (r)$, in an approximation of first order, is the gravitational newtonian potential \cite{wald}. Comparing (\ref{ele}) and (\ref{ele1}), we get $a(r)=2\Phi (r)$ and $e^{-b(r)}=1-(2M/r)$. Substituting this into (\ref{te}), we obtain 
\begin{equation}
T(r)=\frac{2}{r^3}\left[1+2r\Phi^{\prime}(r)\right]\left[r-2M(r)\right]\label{t6}\;.
\end{equation}
Inserting (\ref{t6}) into (\ref{presr}), we get 
\begin{equation}
\frac{d\Phi}{dr}=\frac{M(r)+4\pi r^3 p_{r}}{r[r-2M(r)]}\label{pg}\; ,
\end{equation}
which, for the Newtonian limit $4\pi r^3p_{r}\ll M(r)\ll r$ \cite{wald}, is
\begin{equation}
\frac{d\Phi}{dr}\approx \frac{M(r)}{r^2}\label{pgln}\;.
\end{equation}
In this limit, multiplying (\ref{pgln}) by $r^2$, differentiating, with the account of (\ref{mass}), we obtain the Poisson equation
\begin{equation}\label{pe}
\frac{1}{r^2}\frac{d}{dr}\left(r^2\frac{d\Phi}{dr}\right)=4\pi \rho\;.
\end{equation}
Now, we can obtain the equation of conservation in our case. Deriving (\ref{presr}) with respect to $r$, we get
\begin{equation}
4\pi p_{r}^{\prime}=\frac{T^{\prime}}{4}+\frac{1}{r^3}\label{prp}\;.
\end{equation} 
Summing (\ref{dens}) with (\ref{presr}) and multiplying by $a^{\prime}/2$, we obtain 
\begin{equation}\label{denspr}
2\pi a^{\prime}(\rho+p_{r}^{\prime})=\frac{a^{\prime}e^{-b}}{4r}(a^{\prime}+b^{\prime})\;.
\end{equation}
For the isotropic case 
\begin{eqnarray}
4\pi(p_{t}-p_{r})&=&\frac{e^{-b}}{2}\left[\frac{2a^{\prime}}{r}+\frac{2}{r^2}-a^{\prime\prime}-\frac{\left(a^{\prime}\right)^2}{2}+\frac{a^{\prime}b^{\prime}}{2}-\frac{(a^{\prime}-b^{\prime})}{r}\right]-\frac{1}{r^2}=0\nonumber\\
\Rightarrow a^{\prime\prime}&=&\frac{2a^{\prime}}{r}+\frac{2}{r^2}-\frac{\left(a^{\prime}\right)^2}{2}+\frac{a^{\prime}b^{\prime}}{2}-\frac{(a^{\prime}-b^{\prime})}{r}-\frac{2e^{-b}}{r^2}\label{app}
\end{eqnarray}
Summing (\ref{prp}) with (\ref{denspr}), and taking into account (\ref{app}), we get 
\begin{equation}
4\pi p_{r}^{\prime}+2\pi a^{\prime}\left(\rho+p_{r}\right)=0\label{ec}\;,
\end{equation}
where  $p_{r}=p_{t}$. Considering now that the line element  (\ref{ele1}),  (\ref{ec}) becomes 
\begin{equation}
\frac{dp_{r}}{dr}=-\left(\rho+p_{r}\right)\frac{d\Phi}{dr}\label{p2}\;,
\end{equation}
which, in the Newtonian limit $p_{r}\ll \rho$ \cite{wald}, turns into
\begin{equation}
\frac{d\Phi}{dr}\approx -\frac{1}{\rho}\frac{dp_{r}}{dr}\label{prln}\;.
\end{equation}
Substituting  (\ref{prln}) into  (\ref{pe}), we get 
\begin{equation}
\frac{1}{r^2}\frac{d}{dr}\left(\frac{r^2}{\rho}\frac{dp_{r}}{dr}\right)=-4\pi \rho\label{eppr}\;.
\end{equation}
We assume that we may model the structures as polytropic  distributions, and then \cite{hansen}
\begin{equation}
p_{r}=K\rho^{\gamma}\;,
\end{equation}
where $K,\gamma\in\Re$. Defining the so-called polytropic  index  $n=1/(\gamma-1)$, we may write the energy density as $\rho(r)=\rho_{c}\theta^{n}(r)$, with $0\leqslant\theta (r)\leqslant1$ and $\rho_{c}$ being the energy density at the center, and making use of the coordinate transformation $r=\alpha x$, with $\alpha=(\rho^{(1-n)/(2n)}_{c}/2)\sqrt{(n+1)K/\pi}$, the equation  (\ref{eppr}) becomes the equation of  Lan\'{e}-Emden
\begin{equation}
\frac{1}{x^2}\frac{d}{dx}\left(x^2\frac{d\theta (x)}{dx}\right)=-\theta^{n} (x)\label{eLE}\;.
\end{equation}
Taking the initial conditions $\theta(x_c=0)=1,\theta ^{\prime}(0)=0$, with $x_{c}$ at the center, this equation possesses exact solutions, among which, for  $n=0,1,5$ ($\gamma=+\infty , 2, 6/5$), we get 
\begin{eqnarray}
\left\{\begin{array}{lll}
\theta_{(0)}(x)=1-\frac{x^2}{6}\;,\;x_{s}=\sqrt{6}\;,\\
\theta_{(1)}(x)=\frac{\sin x}{x}\;,\; x_{s}=\pi\;,\\
\theta_{(5)}(x)=\left(1+x^2/3\right)^{-1/2}\;,\;x_{s}\rightarrow\infty\;,
\end{array}\right.
\end{eqnarray}
with $x_s$ being the value at the surface of the star. The function $\theta(x)$ must satisfy the condition $\theta(x_{s})=0$. Taking into account (\ref{eLE}), the radius and the total mass of the star are given by \cite{hansen}
\begin{eqnarray}
r_s&=&\alpha x_s\;,\\
M(r_s)&=&4\pi \alpha^3 \rho_c x_s^2 \frac{d\theta}{dx}\Big|_{x=x_s}\;.
\end{eqnarray}
The well known values of the polytropic index are: a) the relativistic case $n=3$ ($\gamma=4/3$); b) the non-relativistic case $n=3/2$ ($\gamma=5/3$).
\par
If we consider a choice of tetrad as non-diagonal, as in \cite{boehmer,gamal}, and taking the algebraic function $f(T)=T$, the same results shown here are recovered. Since our goal in this work is to deal with diagonal tetrad, we propose to investigate the contribution of the terms of higher orders in the torsion, as $T^m$, with $m\in\Re$, in a future work. The contribution of a general algebraic function in $f(R)$ theory, has been done in \cite{capozziello2}, showing a generalized equation of Lan\'{e}-Emden. We hope that a similar result may be found in the case of non-diagonal tetrad for the $f(T)$ theory.

\section{Conclusion}
We considered the $f(T)$ theory coupled with an anisotropic fluid. For static spacetimes with spherically symmetry, we obtained several classes of solutions by imposing conditions to the metric functions $a(r)$, $b(r)$ and to the matter content, represented by the energy density  $\rho (r)$ and the radial pressure  $p_{r}(r)$  and the tangential $p_{t}(r)$.
\par
We analysed first the general isotropic case ($p_{r} = p_{t}$), which results into the equation (\ref{isog}). This equation is solved for three cases, $T = 0$ , $T\neq0$ with the conditions (\ref{cond1}) and (\ref{cond2}). For $ T = 0 $, we regained the  Boehmer's solution \cite{boehmer}, which in GR is classified in \cite{semiz}. This solution has a singularity at $r = 0$ and can possess a metric with the signature $(+---)$ and $(++--)$, depending on the choice of $ c_{0}$ with respect to $r$. The case of the non vanishing  torsion and  the condition (\ref{cond1}), we obtained a new solution with the matter content singular  at $r=0$. For the case of the condition (\ref{cond2}), the solution obeys an equation of state identical  to that of dark energy and which is singular at $r =0$. The solution (\ref{desol}) is a periodic solution which limits the value of the radial coordinate $r$, according to the choice of the constants $c_{0}$ and $c_{1}$. 
\par
When the matter content is anisotropic, we supposed that the energy density, the radial and tangential pressures depend on the algebraic functions $f(T)$ and $f_{T}(T)$, and also on arbitrary functions of the radial coordinate $r$ according to the  generalized conditions (\ref{gcond1})-(\ref{gcond3}). When the energy density obeys the condition (\ref{gcond1}), we have three new solutions for the generalized coordinate conditions (\ref{cond1}), (\ref{cond2}) and (\ref{cond3}). When the tangential pressure obeys the condition (\ref{gcond3}), we get an explicit example of a new solution to the generalized coordinate condition (\ref{cond1}). When the radial pressure obeys the generalized condition (\ref{gcond2}), this resulted into three new solutions, considering the conditions of coordinate (\ref{cond1}), (\ref{cond2}) and (\ref{cond3}) . These solutions generalize our solutions previously obtained in \cite{stephane}, according to the choice of functions $g_{4}(r)$, $g_{5}(r)$ and $g_{6}(r)$.
\par
We present a short summary of the reconstruction method for a spherically  symmetric and static case for the  $f(T)$ theory. This method has proven effective in the regaining and even of obtaining new solutions. We show two simplest examples in the Section $5$, reconstructing the algebraic function $f(T)$ in (\ref{fr1.1}) and (\ref{fr2.1}), as a linear function in $T$, always obeying the imposition (\ref{impf}), arising from the non-diagonal equation (\ref{impos}). This is a consequence of the choice of a set of diagonal tetrad, for characterizing the frame of the metric (\ref{ele}). As this choice is not unique, we may also introduce other sets of tetrad matrices for the most comprehensive study of the reconstruction of the function $f(T)$, and may not be linear.
\par
Due to our conjecture for the $f(T)$ theory, i.e., the choice of a set of diagonal tetrad, resulting into a linear function for the scalar torsion, the regain the same result as in GR, for the stellar structure. The equation of Lan\'{e}-Emden has been obtained for an approximation of first order.
\par
Through a clear methodology, we found various classes of anisotropic solutions, and three isotropic one, for $f(T)$ theory. We conclude that some conditions on the coordinates, the energy density and pressures, can generate new classes of anisotropic and isotropic solutions. Through new generalized conditions, it can be found a range of new solutions in this theory. We believe that the introduction of new symmetries, as that of Papapetrou in Gravitation \cite{amir},  and that of Lema\^{i}tre-Tolman-Bondi in  Cosmology \cite{tolman}, leads to a range of new interesting solutions for the $f(T)$  theory. We also need to check
the validity of the energy dominant condition, $\rho\geq 0$, for our solutions. We propose to present  this in  forthcoming work. We also hope that the introduction of a set of non-diagonal tetrads for characterizing the line element, we will obtain a modified equation of Lan\'{e}-Emben in the $f(T)$ gravity, for the stellar structure in an approximation of first order. 


\vspace{0,25cm}
{\bf Acknowledgement:}   M. H. Daouda thanks CNPq/TWAS for financial support. M. E. Rodrigues  thanks  UFES for the hospitality during the development of this work. M. J. S. Houndjo thanks  CNPq for partial financial support.

\end{document}